\documentclass[12pt]{elsarticle}

\usepackage[centertags]{amsmath}
\usepackage{amsfonts}
\usepackage{amssymb}
\usepackage{amsthm}
\usepackage{newlfont}
\usepackage{txfonts}
\usepackage{epsfig}
\usepackage{amscd}

\newcommand{\RR}{{\mathbb R}}

\newcommand{\CC}{{\mathbb C}}

\newcommand{\beq}{\begin{equation}}
\newcommand{\eeq}{\end{equation}}
\newcommand{\ba}{\begin{array}}
\newcommand{\ea}{\end{array}}
\newcommand{\bea}{\begin{eqnarray}}
\newcommand{\eea}{\end{eqnarray}}

\newcommand{\sgn}{\mathop{\mathrm{sgn}}}

\usepackage{graphicx}
\usepackage[usenames,dvipsnames]{color}
\usepackage{transparent} 

\numberwithin{equation}{section}
\allowdisplaybreaks

\begin{document}
\title{Nonlocality and the inverse scattering transform for the Pavlov equation\tnoteref{t1}}
\tnotetext[t1]{The first author was partially supported by the Russian Foundation for Basic Research, grant 13-01-12469 ofi-m2, Russian Federation Government grant No~2010-220-01-077, by the program ``Leading scientific schools'' (grant NSh-4833.2014.1), by the program ``Fundamental problems of nonlinear dynamics'', Presidium of RAS, by the INFN sezione di Roma, and by the PRIN 2010/11 No~JJ4KPA\_004 of Roma 3}
\author[lan]{P.G.~Grinevich\corref{cor1}\fnref{fn1}}
\ead{pgg@landau.ac.ru}
\author[ur1]{P.M.~Santini}
\ead{paolo.santini@roma1.infn.it}
\address[lan]{L.D. Landau Institute for Theoretical Physics,
pr. Akademika Semenova 1a, 
Chernogolovka, 142432, Russia; Lomonosov Moscow State University, Faculty of Mechanics and Mathematics, 
Russia, 119991, Moscow, GSP-1, Leninskiye Gory 1, Main Building;
Moscow Institute of Physics and Technology, 9 Institutskiy per., Dolgoprudny,
Moscow Region, 141700, Russia}
\address[ur1]{Dipartimento di Fisica, Universit\`a di Roma ``La Sapienza'', 
Piazzale Aldo Moro 2, I-00185 Roma, Italy; 
Istituto Nazionale di Fisica Nucleare, Sezione di Roma,
Piazzale Aldo Moro 2, I-00185 Roma, Italy}

\begin{abstract}
As in the case of soliton PDEs in 2+1 dimensions, the evolutionary form of integrable dispersionless multidimensional PDEs
is non-local, and the proper choice of integration constants should be the one dictated by the associated 
Inverse Scattering Transform (IST). Using the recently made rigorous IST for vector fields associated with the so-called 
Pavlov equation $v_{xt}+v_{yy}+v_xv_{xy}-v_yv_{xx}=0$, in this paper we establish the following. 1. The non-local term 
$\partial_x^{-1}$ arising from its evolutionary form $v_{t}= v_{x}v_{y}-\partial^{-1}_{x}\,\partial_{y}\,[v_{y}+v^2_{x}]$ corresponds 
to the asymmetric integral $-\int_x^{\infty}dx'$. 2. Smooth and well-localized initial data $v(x,y,0)$ evolve in time developing,
for $t>0$, the constraint $\partial_y {\cal M}(y,t)\equiv 0$, where ${\cal M}(y,t)=\int_{-\infty}^{+\infty} \left[v_{y}(x,y,t) +(v_{x}(x,y,t))^2\right]\,dx$. 3. Since no smooth and well-localized initial data can satisfy such constraint at $t=0$, the initial 
($t=0+$) dynamics of the Pavlov equation can not be smooth, although, as it was already established,  small norm solutions remain regular for all positive times. We expect that the techniques developed in this paper to prove the above results, should be 
successfully used in the study of the non-locality of other basic examples of integrable dispersionless PDEs in 
multidimensions.  
\end{abstract}

\maketitle

\section{Introduction}

Integrable dispersionless PDEs in multidimensions, intensively studied in the recent literature (see \cite{MS5} for an account 
of the vast literature on this subject), arise as the condition of commutation $[L,M]=0$ of pairs of one-parameter families of 
vector fields. A novel Inverse Scattering Transform (IST) for vector fields has been constructed, at a formal level in  
\cite{MS0}, \cite{MS1}, \cite{MS2}, to solve their Cauchy problem, obtain the long-time asymptotics, and establish if, due 
to the lack of dispersion, the nonlinearity is strong enough to cause a gradient catastrophe at finite time. Due to the novel 
features of such IST (the corresponding operators are unbounded, the kernel space is a ring, the inverse problem is 
intrinsically non-linear), together with the lack of explicit regular localized solutions, it was clearly important to make 
this IST rigorous, and this goal was recently achieved in \cite{GSW} on the illustrative example of the so-called Pavlov equation 
\cite{Pavlov}, \cite{Ferapontov}, \cite{Duna}, 
\beq 
\label{Pavlov}
\ba{l}
v_{xt}+v_{yy}+v_xv_{xy}-v_yv_{xx}=0,~~v=v(x,y,t)\in\RR,~~x,y,t\in\RR, 
\ea
\eeq
arising in the study of integrable hydrodynamic chains  \cite{Pavlov}, and in Differential Geometry as a particular example of Einstein - Weyl metric \cite{Duna}. It was first derived in \cite{Duna1} as a conformal symmetry of the second heavenly equation. 

In the form (\ref{Pavlov}) it is not an evolution equation. To rewrite it in the evolution form, we have to integrate 
it with respect to $x$:
 \beq 
\label{Pavlov-ev}
v_{t}= v_{x}v_{y}-\partial^{-1}_{x}\,\partial_{y}\,[v_{y}+v^2_{x}],~~v=v(x,y,t)\in\RR,~~x,y,t\in\RR, 
\eeq
where $\partial^{-1}_{x}$ is the formal inverse of $\partial_{x}$. Of course, it is defined up to an arbitrary integration constant 
depending on $y$ and
$t$. On the other hand, the IST for integrable dispersionless PDEs provides us with a unique solution of the Cauchy problem
in which the function $v(x,y,0)$ is assigned, corresponding to a 
specific choice of such integration constant. The main goal of this paper is to specify the choice of the integration constant in this specific example.

More precisely, we show that the IST formalism corresponds to the following evolutionary form of the Pavlov equation:
 \beq 
\label{Pavlov-ev2}
v_{t}(x,y,t)= v_{x}(x,y,t)\,v_{y}(x,y,t) +\int_{x}^{+\infty} [v_{y}(x',y,t) +(v_{x'}(x',y,t))^2]_y\,dx', \ \ t\ge0. 
\eeq
In addition, for any smooth compact support initial condition and any $t>0$, the solution develops the constraint 
\begin{equation}
\label{eq:man-cond}
\partial_y {\cal M}(y,t)\equiv 0, \ \ \mbox{where} \ \ 
{\cal M}(y,t)=\int_{-\infty}^{+\infty} \left[v_{y}(x,y,t) +(v_{x}(x,y,t))^2\right]\,dx,
\end{equation}
identically in $y$ and $t$, but, unlike the  Manakov constraints for the Kadomtsev-Petviashvili  (KP) \cite{KP} and for the  
dispersionless Kadomtsev-Petviashvili  (dKP) \cite{ZK} equations, no rapidly decaying smooth initial data can satisfy this condition at $t=0$. Indeed, if we have well-localized Cauchy data, then ${\cal M}(y,0)=\mbox{const}$, and ${\cal M}(y,0)\rightarrow0$ for $|y|\rightarrow\infty$; therefore ${\cal M}(y,0)\equiv0$. On the other hand,
$$
\int\limits_{-\infty}^{+\infty} {\cal M}(y,0)\, dy= \int\limits_{-\infty}^{+\infty} \int\limits_{-\infty}^{+\infty}
(v_{x}(x,y,0))^2 \,dxdy >0,
$$
unless $v_{x}(x,y,0)\equiv0$.

We expect that the techniques developed in this paper to prove the above results, should be 
successfully used in the study of the non-locality of other basic examples of integrable dispersionless PDEs in 
multidimensions.

Let us point out that the problem of non-locality is not typical of integrable dispersionless PDEs only, but it is also 
a generic feature of soliton PDEs with 2 spatial variables. Therefore the problem of choosing proper integration constants is very 
important also in the soliton contest, and the IST provides the natural choice.  This problem was first posed and discussed 
in \cite{AW} for the KP equation. The final answer for KP was obtained in \cite{BPP}, and, later, in \cite{FS}. 

It is interesting to remark the following. The analogue of the constraint (\ref{eq:man-cond}) for KP (and dKP) 
\cite{AW}, \cite{BPP}, \cite{FS}
$$
\partial^2_y \int_{-\infty}^{+\infty} u(x,y,t)\,dx\equiv 0
$$
can be easily satisfied for a large class of regular well-localized initial data and, for such data, the 
initial ($t=0+$) dynamics is smooth but, for the dKP equation it typically results in singularities at finite time \cite{MS4}. On 
the contrary, for Pavlov equation, no smooth well-localized initial data can be chosen to have a smooth initial dynamics, but 
small-norm solutions remain regular for all positive times \cite{GSW}.

In the remaining part of this introduction we summarize the basic formulas of the IST for the Pavlov equation (see, for 
instance, \cite{GSW}) that will be used in this paper.

\subsection{Summary of the IST for the Pavlov equation}

The Pavlov equation is the commutativity condition $[L,M]=0$ for the following pair of vector fields: 
\begin{align}
\label{Lax_Pavlov}
L\equiv \partial_y+(\lambda +v_x)\partial_x, \\
M\equiv \partial_t+(\lambda^2+\lambda v_x-v_y)\partial_x.\nonumber
\end{align}
Assuming, as in \cite{GSW}, that the smooth Cauchy datum $v(x,y,0)$ has compact support, we define the spectral data using the following procedure: 
\begin{enumerate}
\item We define the real Jost eigenfunctions $\varphi_{\pm}(x,y,\lambda)$, $\lambda\in\RR$ as the solutions of the equation
$$
L\varphi_{\pm}(x,y,\lambda)=0,
$$
with the boundary condition:
$$
\varphi_{\pm}(x,y,\lambda)\rightarrow x-\lambda y \ \ \mbox{as} \ \ y\rightarrow\pm\infty,
$$
using the correspondent vector fields ODE:
\begin{equation}
\label{ODE}
\frac{dx}{dy}=\lambda+v_x(x,y)
\end{equation}
\item If we denote by $x_{-}(y,\tau,\lambda)$ the solution of (\ref{ODE}) with the following asymptotics:
$$
x_{-}(y,\tau,\lambda)=\tau+\lambda y + o(1) \ \ \mbox{as} \ \ y\rightarrow -\infty,
$$
then the classical time-scattering datum $\sigma(\tau,\lambda)$ is defined through the following formula:
\begin{equation}
\label{def_sigma}
\sigma(\tau,\lambda)=\lim\limits_{y\rightarrow+\infty} [x_{-}(y,\tau,\lambda)-\tau-\lambda y].
\end{equation}
Equivalently, 
$$
\sigma(\tau,\lambda)=\int_{-\infty}^{\infty} v_x (x_{-}(y,\tau,\lambda),y)dy.
$$
In the linear limit $|v|\ll 1$, the scattering datum $\sigma(\tau,\lambda)$ reduces to the Radon transform 
of $v_x(x,y)$ \cite{MS3}.
\item 
The spectral data $\chi_{\pm}(\tau,\lambda)$ are defined as the solutions of the following shifted Riemann-Hilbert (RH)  problem: 
\beq\label{E:shift-intro}
\sigma(\tau,\lambda)+\chi_{+}(\tau+\sigma(\tau,\lambda),\lambda)-\chi_{-}(\tau,\lambda)=0,\quad \tau,\lambda \in \RR,
\eeq
where
$\chi_{\pm}(\tau,\lambda)$ are analytic in $\tau$ in the upper and lower half-planes $\CC^{\pm}$  respectively, and
$$
\chi_{\pm}(\tau,\lambda)\to 0\ \ \mbox{as}\ \ |\tau|\to\infty.
$$
\end{enumerate}

If the potential $v(x,y,t)$ evolves in $t$ according to the Pavlov equation, then the scattering and the spectral data evolve in a simple way:
\begin{align}
\label{eq:spectral-evolution}
&\sigma(\tau,\lambda,t)= \sigma(\tau-\lambda^2 t,\lambda,0),\\
&\chi_{\pm}(\tau,\lambda,t)= \chi_{\pm}(\tau-\lambda^2 t,\lambda,0).\nonumber
\end{align}
 
The reconstruction of the solution consists of two steps:
\begin{enumerate}
\item
One solves the following nonlinear integral equation for the time-dependent real Jost eigenfunction:
\beq
\label{inversion1}
\psi_-(x,y,t,\lambda)-H_{\lambda}\chi_{-I}\big(\psi_-(x,y,t,\lambda),\lambda\big)+\chi_{-R}\big(\psi_-(x,y,t,\lambda),\lambda\big)=x-\lambda y-\lambda^2 t,
\eeq
where $\chi_{-R}$ and $\chi_{-I}$ are the real and imaginary parts of $\chi_-$ , and $H_{\lambda}$ is the Hilbert transform operator wrt $\lambda$
\beq
H_{\lambda}f(\lambda)=\frac{1}{\pi} \fint\limits_{-\infty}^{\infty}\frac{f(\lambda')}{\lambda-\lambda'}d\lambda' .
\eeq
In \cite{GSW} it is shown that, for Cauchy data satisfying some explicit small-norm conditions, 
equation (\ref{inversion1}) is uniquely solvable for all $t\ge0$. 
\item Once the real time-dependent Jost eigenfunction is known, the solution $v(x,y,t)$ of the Pavlov equation is defined 
by:
\begin{equation}
\label{eq:v-riemann}
v(x,y,t)= -\frac{1}{\pi} \int\limits_{\RR} \chi_{-I}(\psi_{-}(x,y,t,\zeta),\zeta) d\zeta.
\end{equation}
\end{enumerate}

In addition, in \cite{GSW} it was shown that, under the same analytic assumptions on the Cauchy data, 
the function $\omega(x,y,t,\lambda)=\psi_-(x,y,t,\lambda)-x+\lambda y + \lambda^2 t$ belongs to 
the spaces $L^{\infty}(d\lambda)$ and $L^{2}(d\lambda)$ for all real $x$, $y$ and $t\ge0$, and continuously 
depends on these variables. Moreover, for all $x,y\in\RR$, $t\ge 0$, the following derivatives of 
$\omega$: 
$$
\partial_x \omega, \ \ \partial_y \omega, \ \ \partial_t \omega,  \ \ \partial^2_x \omega, 
\ \ \partial^2_y \omega, \ \ \partial_x\partial_y \omega, \ \ \partial_t\partial_x \omega,
$$
are well-defined as elements of the space $L^2(d\lambda)$, they 
continuously depend on $x,y,t$ and are uniformly bounded in $\RR\times\RR\times \overline{\RR^+}$.

For $t>0$ we have
\begin{equation}
v_{t}(x,y,t)=-\frac 1{\pi}\int_{\mathbb R}\partial_{\tau}\chi_{-I}(x-\lambda y -\lambda^2 t+ \omega(x,y,t,\lambda),
\lambda) \psi_{t} d\lambda,
\end{equation}
but for $t=0$ this integral diverges and the calculation of $v_t$ requires an additional investigation presented below.

Our strategy is the following:
\begin{enumerate}
\item In Section~\ref{sect:lo} we calculate the $t$-derivative of $v(x,y,t)$ for $t\ge 0$ in what we call the ``leading order approximation''.
\item  In Section~\ref{sect:corr} we show that the correction to the leading order approximation vanishes for $x\rightarrow\pm\infty$.
\end{enumerate}

\section{The leading order approximation}
\label{sect:lo}

Let us calculate the $t$-derivative of the function $v(x,y,t)$ at $t=0$ in the leading order approximation. We assume 
the following:
\begin{enumerate}
\item We replace $\sigma(\tau,\lambda)$ by the leading term  $\sigma_{L}(\tau,\lambda)$ (see (\ref{def_sigmaL})) of the $\frac{1}{\lambda}$ expansion corresponding to
$\lambda\rightarrow\pm\infty$.
\item Instead of the shifted RH problem  (\ref{E:shift-intro}) we use the standard RH problem
\beq\label{E:shift-L}
\chi_{L-}(\tau,\lambda)-\chi_{L+}(\tau,\lambda) = \sigma_{L}(\tau,\lambda), \quad \tau\in \RR,
\eeq
where
$\chi_{L\pm}(\tau,\lambda)$ are analytic in $\tau$ in the upper and lower half-planes $\CC^{\pm}$  respectively.
\item In the formula (\ref{eq:v-riemann}) we neglect the $\omega(x,y,t)$ influence, and we write
\begin{equation}
\label{eq:v-riemann-1}
v_{L}(x,y,t)= -\frac{1}{\pi} \int\limits_{\RR} \chi_{L-I}(x-\lambda y -\lambda ^2 t,\lambda) d\lambda.
\end{equation}
\end{enumerate}
\subsection{The leading order of the scattering data}
For calculation of $\sigma(\tau,\lambda)$ at large $|\lambda|$ it is convenient to use $x$ as independent variable and 
$y$ as the dependent one. Then equation (\ref{ODE}) reads as:
\begin{equation}
\label{ODE2}
\frac{dy}{dx}= \frac{1}{\lambda+ v_x(x,y)}= \frac{1}{\lambda}- 
\frac{1}{\lambda^2}v_x(x,y)+\frac{1}{\lambda^3}v^2_x(x,y)+
O\left(\frac{1}{\lambda^4} \right),
\end{equation}
and the boundary condition takes the form:
$$
y=-\tilde\tau+ \frac{x}{\lambda} \ \ \mbox{as} \ \ x\rightarrow-\sgn(\lambda)\cdot\infty
$$
where 
$$
\tilde\tau=\frac{\tau}{\lambda}
$$
Let
$$
y(x) = -\tilde\tau +  \frac{x}{\lambda} + \frac{y_2}{\lambda^2} + O\left(\frac{1}{\lambda^3} \right).
$$
Substituting it into (\ref{ODE2}) we obtain: 
$$
 \frac{1}{\lambda} + \frac{(y_2)_x}{\lambda^2} +
O\left(\frac{1}{\lambda^3} \right)= \frac{1}{\lambda}- 
\frac{1}{\lambda^2}v_x\left(x,-\tilde\tau+ \frac{x}{\lambda} \right)+ O\left(\frac{1}{\lambda^3} \right),
$$
therfore
$$
(y_2)_x= - v_x(x,-\tilde\tau), \ \ y_2(x;\tau)=-v(x,-\tilde\tau),
$$
and
$$
y(x;\tau,\lambda)=-\tilde\tau+\frac{x}{\lambda} -\frac{v(x,-\tilde\tau)}{\lambda^2}+ O\left(\frac{1}{\lambda^3} \right).
$$
Therefore
$$
\sigma(\tau,\lambda)=\sgn(\lambda)\int\limits_{-\infty}^{\infty}v_x(x,y(x;\tau,\lambda))\frac{dy}{dx}dx=
$$
$$
=\sgn(\lambda)\int\limits_{-\infty}^{\infty} v_x\left(x,-\tilde\tau+\frac{x}{\lambda}
-\frac{v(x,-\tilde\tau)}{\lambda^2} \right) \left[\frac{1}{\lambda}- 
\frac{1}{\lambda^2}v_x\left(x,-\tilde\tau+ \frac{x}{\lambda}\right)+\frac{1}{\lambda^3}v^2_x(x,-\tilde\tau)\right]+
O\left(\frac{1}{\lambda^4} \right)=
$$
$$
=\sgn(\lambda)\int\limits_{-\infty}^{\infty}\left[ v_x(x,-\tilde\tau) +\frac{1}{\lambda}
x v_{xy} (x,-\tilde\tau) + \frac{1}{\lambda^2}\left( \frac{x^2}{2}v_{xyy} (x,-\tilde\tau) 
- (vv_{xy})(x,-\tilde\tau) \right)  \right]\times
$$
$$
\times  \left[\frac{1}{\lambda}- 
\frac{1}{\lambda^2}v_x(x,-\tilde\tau)+\frac{1}{\lambda^3}(-xv_{xy}+v^2_x)(x,-\tilde\tau)\right]
+O\left(\frac{1}{\lambda^4} \right)=
$$
$$
=\sgn(\lambda)\int\limits_{-\infty}^{\infty} \left[\frac{1}{\lambda}v_x(x,-\tilde\tau)+
\frac{1}{\lambda^2}(xv_{xy}-v_x^2)(x,-\tilde\tau)+
\right.
$$
$$
\left. +\frac{1}{\lambda^3}\left( \frac{x^2}{2}v_{xyy} - vv_{xy} -2xv_xv_{xy}
+v_x^3   \right) (x,-\tilde\tau)  
\right] dx+O\left(\frac{1}{\lambda^4} \right).
$$
Denoting by:
$$
V_1(y)= \int\limits_{-\infty}^{\infty} v_x(x,y)dx=0,
$$
$$
V_2(y)= \int\limits_{-\infty}^{\infty} (xv_{xy}-v_x^2)(x,y)dx = 
\int\limits_{-\infty}^{\infty} (-v_{y}-v_x^2)(x,y)dx,
$$
$$
V_3(y)= \int\limits_{-\infty}^{\infty} (xv_{xy}-v_x^2)(x,y)dx = 
\int\limits_{-\infty}^{\infty} \left(\frac{x^2}{2}v_{xyy} - vv_{xy} -2xv_xv_{xy}
+v_x^3 \right)(x,y)dx,
$$
we obtain the following expansion:
$$
\sigma(\tau,\lambda)=
\frac{\sgn(\lambda)}{\lambda^2} V_2\left(-\frac{\tau}{\lambda}\right)+
\frac{\sgn(\lambda)}{\lambda^3} V_3\left(-\frac{\tau}{\lambda}\right)+
O\left(\frac{1}{\lambda^4}\right),
$$
whose leading term reads:
\begin{equation}
\label{def_sigmaL}
\sigma_{L}(\tau,\lambda)=
\frac{\sgn(\lambda)}{\lambda^2} V_2\left(-\frac{\tau}{\lambda}\right).
\end{equation}
\subsection{The leading order of the spectral data}
Let us denote:
$$
\chi_{2-}(\zeta)-\chi_{2+}(\zeta)=V_2(\zeta), \ \ \chi_{2-}(-\zeta)-\chi_{2+}(-\zeta)= V(-\zeta).
$$
Taking into account that $\chi_{2-}(\zeta)$ and $-\chi_{2+}(-\zeta)$ are holomorphic in the lower half-plane, we obtain:
\begin{align*}
&\chi_{L-I}(\tau,\lambda)=\frac{\sgn(\lambda)}{\lambda^2}\, \left[-\chi_{2+I}\left(-\frac{\tau}{\lambda} \right)\right],& 
&\lambda>0,&\\
&\chi_{L-I}(\tau,\lambda)=\frac{\sgn(\lambda)}{\lambda^2}\, \left[\chi_{2-I}\left(-\frac{\tau}{\lambda} \right)\right],& 
&\lambda<0.&
\end{align*}
The function $V_2(\zeta)$ is real; therefore $\chi_{2-I}(\zeta)=\chi_{2+I}(\zeta)$, and
\begin{equation}
\label{def_chiL}
\chi_{L-I}(\tau,\lambda)=-\frac{\chi_{2-I}\left(-\frac{\tau}{\lambda} \right)}{\lambda^2}.
\end{equation} 
\subsection{The leading order approximation for the potential}
From (\ref{def_chiL}) we immediately obtain:
\begin{equation}
v_L(x,y,t)=\frac 1{\pi}\int_{\mathbb R} \frac{\chi_{2-I}\left(-\frac{x}{\lambda}+y+t\lambda \right)}{\lambda^2}   d\lambda.\label{eq:2}
\end{equation}
At $\zeta=\pm\infty$ we have: $\chi_{2-I}(\zeta)=\frac{c_1}{\zeta}+ O(\zeta^{-2})$, where 
$c_1=-\frac{1}{2\pi}\int_{\RR} V_2(y)\,dy$; therefore this integral is well-defined in the sense of principal value.
\subsection{The time-derivative in the leading order approximation}
Assume that $t\ge 0$,  $|\Delta t|\ll 1$ and $\Delta t>0$. Let us calculate the leading order of $v_L(x,y,t+\Delta t)-v_L(x,y,t)$.

It is convenient to introduce the new variable:
$$
z=\frac{\tau}{\lambda}+y
$$
On the line $\tau=x-\lambda y$ we have:
$$
\lambda=\frac{x}{z}, \ \ \left|\frac{d\lambda}{\lambda^2} \right|= \left|\frac{1}{x}\right||dz|.
$$
\begin{equation}
v_L(x,y,0)=\frac 1{|x|\pi}\int_{\mathbb R} \chi_{2-I}\left(y-z \right)   dz,\label{eq:3}
\end{equation}
The straight line $\tau=\lambda z- \lambda y$ and the parabola $\tau=x-\lambda y- t\lambda^2$ intersects at the points:
$$
\lambda_{1,2}=\frac{1}{2t}\left(\pm\sqrt{z^2+4tx}-z\right),
$$
and 
$$
\frac{1}{\lambda_{i}(z)}\frac{\partial\lambda_i(z)}{\partial z}=\mp\frac{1}{\sqrt{z^2+4tx}}, \ \ 
i=1,2.
$$
The next step depends on the sign of $x$.
\begin{enumerate}
\item Let $x>0$. Then $\lambda_1(z)>0$, $\lambda_2(z)<0$, and   
$$
\left|\frac{1}{\lambda_1^2(z)} \frac{\partial\lambda_1(z)}{\partial z} \right|+\left|\frac{1}{\lambda_2^2(z)} \frac{\partial\lambda_2(z)}{\partial z} \right| = -\frac{1}{\lambda_1^2} \frac{\partial\lambda_1}{\partial z} -\frac{1}{\lambda_2^2} \frac{\partial\lambda_2}{\partial z} =
$$
$$
=\frac{2t}{\left(\sqrt{z^2+4tx}-z\right) \left(\sqrt{z^2+4tx}\right)}- \frac{2t}{\left(-\sqrt{z^2+4tx}-z\right) \left(\sqrt{z^2+4tx}\right)} = \frac{1}{x},
$$
therefore
$$
v_L(x,y,t)=\frac 1{\pi}\int\limits_{\mathbb R} \chi_{2-I}\left(y-z \right)\left[
\left|\frac{1}{\lambda_1^2} \frac{\partial\lambda_1}{\partial z} \right|+\left|\frac{1}{\lambda_2^2} \frac{\partial\lambda_2}{\partial z} \right|\right] dz =\frac 1{|x|\pi}\int\limits_{\mathbb R} \chi_{2-I}\left(y-z \right) dz= v_L(x,y,0),
$$
and
$$
\partial_t v_{L}(x,y,t)=0, \ \ t\ge 0.
$$
\item Let $x<0$. Then $\sgn{\lambda_1(z)}=\sgn{\lambda_2(z)}=-\sgn{z} $, and
$$
\left|\frac{1}{\lambda_1^2} \frac{\partial\lambda_1}{\partial z} \right|+\left|\frac{1}{\lambda_2^2} \frac{\partial\lambda_2}{\partial z} \right| = \sgn{z}\left[\frac{1}{\lambda_1^2} \frac{\partial\lambda_1}{\partial z} -\frac{1}{\lambda_2^2} \frac{\partial\lambda_2}{\partial z}\right] =
$$
$$
=\sgn{z}\left[\frac{2t}{\left(z-\sqrt{z^2+4tx}\right) \left(\sqrt{z^2+4tx}\right)}+ \frac{2t}{\left(z+\sqrt{z^2+4tx}\right) \left(\sqrt{z^2+4tx}\right)}\right] =
$$
$$
=\frac{2t\sgn{z}}{\sqrt{z^2+4tx}}\cdot\frac{2z}{z^2-z^2-4tx}=\frac{1}{-x}\frac{|z|}{\sqrt{z^2+4tx}}=\frac{1}{|x|} \frac {1}{\sqrt {1+ \frac {4 t x}{z^2}}}.
$$
Taking into account that, for $x<0$, the variable $z$ runs through the intervals $|z|\ge2\sqrt{t|x|}$, we obtain:
$$
v_L(x,y,t)=\frac 1{|x|\pi}\left[\left(\int\limits_{-\infty}^{-2\sqrt{t|x|}} + \int\limits_{2\sqrt{t|x|}}^{\infty} \right)
\left(
\chi_{2-I}\left(y-z \right)\left[
\left|\frac{1}{\lambda_1^2} \frac{\partial\lambda_1}{\partial z} \right|+\left|\frac{1}{\lambda_2^2} \frac{\partial\lambda_2}{\partial z} \right|\right] dz\right) \right]=
$$
$$
=\frac 1{|x|\pi}\int\limits_{2\sqrt{t|x|}}^{\infty}
\frac{\chi_{2-I}\left(y-z \right)+ \chi_{2-I}\left(y+z \right)}{\sqrt {1+ \frac {4 t x}{z^2}}}\,dz.
$$
$$
\partial_tv_L(x,y,t)=\frac 1{|x|\pi} \lim\limits_{\alpha\rightarrow0+}
\partial_t\!\!\!\!\! \int\limits_{\sqrt{4t|x|+\alpha}}^{\infty} 
\frac{[\chi_{2-I}(y+ z) + \chi_{2-I}(y-z)]\,z dz}{\sqrt {z^2- 4 t |x|}}=
$$
$$
=\frac 1{|x|\pi} \lim\limits_{\alpha\rightarrow0+}\left[-\frac{2|x|}{\sqrt{4t|x|+\alpha}}
\frac{\left[\chi_{2-I}(y+\sqrt{4t|x|+\alpha}) + \chi_{2-I}(y-\sqrt{4t|x|+\alpha})\right]\,\sqrt{4t|x|+\alpha}}{\sqrt {\alpha}}+\right.
$$
$$
\left.+\int\limits_{\sqrt{4t|x|+\alpha}}^{\infty} 
\frac{2|x|[\chi_{2-I}(y+ z) + \chi_{2-I}(y-z)]\,z dz}{(z^2- 4 t |x|)^{3/2}}\right]=
$$
$$
=\frac 1{|x|\pi} \lim\limits_{\alpha\rightarrow0+}\left[-\frac{2|x|}{\sqrt{\alpha}}
\left[\chi_{2-I}(y+\sqrt{4t|x|+\alpha}) + \chi_{2-I}(y-\sqrt{4t|x|+\alpha})\right]-\right.
$$
$$
\left.-\int\limits_{\sqrt{4t|x|+\alpha}}^{\infty}2|x| 
\left(\frac{1}{\sqrt{z^2- 4 t |x|}}\right)_z\cdot [\chi_{2-I}(y+ z) + \chi_{2-I}(y-z)]dz    \right]=
$$
$$
=\frac 2{\pi}\int\limits_{\sqrt{4t|x|}}^{\infty}
\frac{[\chi'_{2-I}(y+ z) - \chi'_{2-I}(y-z)]dz}{\sqrt{z^2- 4 t |x| } }.
$$
Therefore, for a fixed negative $x$ and $t\rightarrow0+$, we obtain:
$$
\partial_t v_{L}(x,y,t)\big|_{t=0+}= \frac{2}{\pi} \fint\limits_{-\infty}^{\infty} \frac{\chi'_{2-I}(y+z)}{z}dz=
\frac{2}{\pi} \fint\limits_{-\infty}^{\infty} \frac{\chi'_{2-I}(z)}{z-y}dz=-2H_y\cdot\chi'_{2-I}(y).
$$
Let us recall that
$$
\chi'_{2-I}(y)=-\frac{1}{2}H_y\cdot V'_2(y), \  \  \mbox{and} \ \ V'_2(y) =2H_y\cdot\chi'_{2-I}(y)
$$
therefore:
$$
\partial_t v_{L}(x,y,t)\big|_{t=0+}=-\partial_yV_2(y)=\int_{-\infty}^{+\infty} [v_{y}(x',y,0) +(v_{x'}(x',y,0))^2]_y\,dx'.
$$
We see that, for $t=0+$ and $x<0$, the function $\partial_t v_{L}(x,y,t)$ does not depend on $x$. On the contrary, if $t>0$, 
then this function decays at $x\rightarrow-\infty$ as $O\left(\frac{1}{(t|x|)^{3/2}}\right)$. 
\end{enumerate}

\section{Corrections to the leading order approximation for large $x$}
\label{sect:corr}
Let us show that, for $|x|\rightarrow\infty$, the exact formulas are well approximated by the leading order formula. 
More precisely, we show the following: the $\lambda$-integration line can be split into two parts. In the first 
part the leading term approximation does not work, but the relative size of this part is small. In the remaining part 
the leading term gives the main influence to the answer. 

Using the estimates from the paper \cite{GSW}, one can easily show that there exists a constant $C$ such that:
\begin{equation}
\label{eq:3.1}
|\partial_{\lambda}\chi_{-}(\tau,\lambda)|\le \frac{C}{1+|\lambda|}, \ \ \ 
|\partial^k_{\tau}\chi_{-}(\tau,\lambda)|\le  \frac{C}{1+|\lambda|^{2+k}}, \ \ k=0,1,2,
\end{equation}
and in the area $|\tau|>2 (D_x+|\lambda|D_y)$, where $-D_x\le x\le D_x$, $-D_y\le y\le D_y$ is the rectangular box containing the Cauchy data support, one has:
\begin{equation}
\label{eq:3.2}
|\partial^k_{\tau}\chi_{-}(\tau,\lambda)|\le \frac{C}{(1+|\lambda|)|\tau|^{k+1}}, \ \ k=0,1,2.
\end{equation}
Using the estimates (\ref{eq:3.1}), (\ref{eq:3.2}) and the arguments from the paper \cite{GSW} used in the proof of Theorem~4.3, one can easily derive that, for a compact area in the $(y,t)$-plane and large $|x|$, the following estimates are valid: 
\begin{equation}
\label{eq:3.3}
\|\omega \|_{L^{\infty}(d\lambda)}=O\left(\frac{1}{|x|^{3/4}}\right). \ \
\|\omega \|_{L^{2}(d\lambda)}=O\left(\frac{1}{|x|^{3/4}}\right), \ \
\|\omega_t \|_{L^{2}(d\lambda)}=O\left(\frac{1}{|x|^{1/4}}\right).
\end{equation}
As a corollary,  $\|\omega \|_{L^{\infty}(d\lambda)}$, $\|\omega \|_{L^{2}(d\lambda)}$ and $\|\omega_t \|_{L^{2}(d\lambda)}$ are 
uniformly bounded in $x$ for any compact area in the $(y,t)$-half-plane.

Let us start from the exact formula. Denote by 
\begin{align}
\label{eq:4}
& v(x,y,t+\Delta t)-v(x,y,t) = \\
&= -\frac{1}{\pi} \int\limits_{\RR} 
\left[\chi_{-I}(x-\lambda y- \lambda^2 t -\lambda^2 \Delta t + \omega(x,y,t+\Delta t,\lambda),\lambda)-
\chi_{-I}(x-\lambda y- \lambda^2 t+ \omega(x,y,t,\lambda),\lambda)\right]   d\lambda.\nonumber
\end{align}
Assume that $|x|$ is large but fixed and $1/(\Delta t)^{3/4}\gg|x|$. Let us denote:
$$
J_1=-\frac 1{\pi}\!\!\!\int\limits_{-\frac{1}{(\Delta t)^{3/4}}}^{\frac{1}{(\Delta t)^{3/4}}} \left[ \chi_{-I}(x-\lambda y-\lambda^2t 
-\lambda^2 \Delta t +\omega(x,y,t+\Delta t,\lambda),\lambda) - \chi_{-I}(x-\lambda y- \lambda^2 t+\omega(x,y,t,\lambda),\lambda) \right]d\lambda,
$$
$$
J_2=-\frac 1{\pi}\int\limits_{|\lambda|\ge\frac{1}{(\Delta t)^{3/4}}} \left[ \chi_{-I}(x-\lambda y-\lambda^2t 
-\lambda^2 \Delta t +\omega(x,y,t+\Delta t,\lambda),\lambda) - \chi_{-I}(x-\lambda y- \lambda^2 t+\omega(x,y,t,\lambda),\lambda) \right]d\lambda,
$$

Let us show that $J_1=-\frac{\Delta t}{\pi} J_1^{(1)}+o(\Delta t)$, where
$$
J_{1}^{(1)}=\int\limits_{-\frac{1}{(\Delta t)^{3/4}}}^{\frac{1}{(\Delta t)^{3/4}}} 
\partial_{\tau}\chi_{-I}(x-\lambda y - \lambda^2 t+ \omega(x,y,t,\lambda),\lambda)
(-\lambda^2 +\omega_{t}(x,y,t,\lambda)) d\lambda.
$$
Using the mean value theorem one can write:
$$
J_1=-\frac 1{\pi}\int\limits_{-\frac{1}{(\Delta t)^{3/4}}}^{\frac{1}{(\Delta t)^{3/4}}} 
\partial_{\tau}\chi_{-I}(\tau_*(x,y,t,\Delta t,\lambda),\lambda)\,
[-\lambda^2\Delta t+\omega(x,y,t+\Delta t,\lambda)-\omega(x,y,t,\lambda)] d\lambda,
$$
The function $\partial_{\tau}\chi_{-I}(\tau_*(x,y,t,\Delta t,\lambda),\lambda)$ is a uniformly bounded in $t$ element of 
$L^2(d\lambda)$, $\omega(x,y,t,\lambda)$ is a differentiable function of $t$ in the space $L^2(d\lambda)$; therefore
$$
J_1=-\frac {\Delta t}{\pi}\int\limits_{-\frac{1}{(\Delta t)^{3/4}}}^{\frac{1}{(\Delta t)^{3/4}}} 
\partial_{\tau}\chi_{-I}(\tau_*(x,y,t,\Delta t,\lambda),\lambda)\,
[-\lambda^2+\omega_t(x,y,t,\lambda)] d\lambda + o(\Delta t),
$$
We also know that
$$
|\tau_*(x,y,t,\Delta t,\lambda)-x+\lambda y+\lambda^2 t-\omega(x,y,t,\lambda)|\le \lambda^2 \Delta t + |\omega(x,y,t+\Delta t,\lambda)-\omega(x,y,t,\lambda)|,
$$
and $\|\omega(x,y,t+\Delta t,\lambda)-\omega(x,y,t,\lambda)\|_{L^2(d\lambda)}=O(\Delta t)$; therefore 
$\left|-\frac{\pi}{\Delta t}J_1-J_1^{(1)}\right|$ is, up to $o(1)$ terms:
$$
\left|\int\limits_{-\frac{1}{(\Delta t)^{3/4}}}^{\frac{1}{(\Delta t)^{3/4}}} [\partial_{\tau}\chi_{-I}(\tau_*(x,y,t,\Delta t,\lambda),\lambda)-\partial_{\tau}\chi_{-I}(x-\lambda y- \lambda^2 t+\omega(x,y,t,\lambda),\lambda)]\, [-\lambda^2+\omega_t(x,y,t,\lambda)]d\lambda\right|\le
$$
$$
\le \int\limits_{-\frac{1}{(\Delta t)^{3/4}}}^{\frac{1}{(\Delta t)^{3/4}}} 
|\partial_{\tau\tau}\chi_{-I}(\tau_{**}(x,y,t,\Delta t,\lambda),\lambda)|\, [\lambda^2 \Delta t + 
|\omega(x,y,t+\Delta t,\lambda)-\omega(x,y,t,\lambda)| ]\,[\lambda^2+|\omega_t(x,y,t,\lambda)|]d\lambda\le
$$
$$
\le \int\limits_{-\frac{1}{(\Delta t)^{3/4}}}^{\frac{1}{(\Delta t)^{3/4}}} \frac{C}{1+\lambda^4}\, [\lambda^2 \Delta t + |\omega(x,y,t+\Delta t,\lambda)-\omega(x,y,t,\lambda)| ]\,[\lambda^2+|\omega_t(x,y,t,\lambda)|] d\lambda\le
$$
$$
\le O\left(\frac{\Delta t}{(\Delta t)^{3/4}}\right)+ \Delta t\int\limits_{-\frac{1}{(\Delta t)^{3/4}}}^{\frac{1}{(\Delta t)^{3/4}}}  \frac{C\lambda^2}{1+\lambda^4} |\omega_t(x,y,t,\lambda)|d\lambda+ \int\limits_{-\frac{1}{(\Delta t)^{3/4}}}^{\frac{1}{(\Delta t)^{3/4}}}  \frac{C\lambda^2}{1+\lambda^4} |\omega(x,y,t+\Delta t,\lambda)-\omega(x,y,t,\lambda)|d\lambda + 
$$
$$
+\int\limits_{-\frac{1}{(\Delta t)^{3/4}}}^{\frac{1}{(\Delta t)^{3/4}}}  |\omega(x,y,t+\Delta t,\lambda)-\omega(x,y,t,\lambda)| |\omega_t(x,y,t,\lambda)|d\lambda=o(1).
$$
Here we used the H\"older inequality for the $L^2(d\lambda)$ functions $\frac{C\lambda^2}{1+\lambda^4}$, 
$\omega(x,y,t+\Delta t,\lambda)-\omega(x,y,t,\lambda)$, and $\omega_t(x,y,t,\lambda)$.

Therefore, for any compact area in the $(y,t)$-variables, we can use $-\frac{\Delta t}{\pi}J_1^{(1)}$ instead of $J_1$ in our calculations of the $t$-derivative.

We have the following corrections to the exact formula in comparison with the leading term approximation:
\begin{enumerate}
\item To express the spectral data $\chi_{-}(\tau,\lambda)$ in terms of the scattering data $\sigma(\tau,\lambda)$  
we use the shifted Riemann-Hilbert problem (\ref{E:shift-intro}) instead of the local one (\ref{E:shift-L}).
\item We neglect the $O\left(\frac{1}{\lambda^3} \right)$ corrections to the leading order approximation for 
the spectral data $\sigma(\tau,\lambda)$. 
\item We replace the function $\omega(x,y,t,\lambda)$ by zero.
\end{enumerate}
Let us estimate these corrections step by step. 

The shifted Riemann-Hilbert problem 
$$
\sigma(\tau,\lambda)+\chi_{+}(\tau+\sigma(\tau,\lambda),\lambda)-\chi_{-}(\tau,\lambda)=0
$$
is equivalent to:
$$
\chi_{-}(\tau,\lambda)-\chi_{+}(\tau,\lambda)=
\sigma(\tau,\lambda)+\chi_{+}(\tau+\sigma(\tau,\lambda),\lambda)-\chi_{+}(\tau,\lambda).
$$
Since 
$$
\chi_{+}(\tau+\sigma(\tau,\lambda),\lambda)-\chi_{+}(\tau,\lambda)=O\left(\frac{1}{|\lambda|^5}\right),
$$
the shifted Riemann-Hilbert problem for $\lambda\rightarrow\infty$ can be approximated 
by the non-shifted one. Taking into account that the next correction to $\sigma(\tau,\lambda)$
has order $O\left(\frac{1}{|\lambda|^3}\right)$, we obtain
\begin{equation}
\label{eq:5}
|\chi_{-}(\tau,\lambda)-\chi_{L-}(\tau,\lambda)|\le \frac{\cal{C}}{1+|\lambda|^3}, \ \ 
| \partial_{\tau}\chi_{-}(\tau,\lambda)- \partial_{\tau}\chi_{L-}(\tau,\lambda)|\le \frac{\cal{C}}{1+|\lambda|^4}.
\end{equation}
Denote by $J_2^{(1)}$ the result of replacing $\chi(\tau,\lambda)$ by $\chi_{L}(\tau,\lambda)$ in $J_2$:
\begin{align*}
J_2^{(1)}=-\frac 1{\pi}\int\limits_{|\lambda|>\frac{1}{(\Delta t)^{3/4}}} &\left[ \chi_{L-I}(x-\lambda y -\lambda^2t -\lambda^2 \Delta t +\omega(x,y,t+\Delta t,\lambda),\lambda) - \right. \\
& -\left. \chi_{L-I}(x-\lambda y -\lambda^2t +\omega(x,y,t,\lambda),\lambda) \right]d\lambda.
\end{align*}
From (\ref{eq:5}) it follows immediately that $J_2-J_2^{(1)}=O((\Delta t)^{3/2})$; therefore, for the calculation of the 
$t$-derivative, one can use  $J_2^{(1)}$ instead of $J_2$. 

Let us now estimate the corrections to $J_1^{(1)}$. It is convenient to split:
$$
J_1^{(1)}=J_{11}+J_{12}
$$
$$
J_{11}=\int\limits_{-\alpha\sqrt{|x|}}^{\alpha\sqrt{|x|}} 
\partial_{\tau}\chi_{-I}(x-\lambda y -\lambda^2 t + \omega(x,y,t,\lambda),\lambda)
(-\lambda^2 +\omega_{t}(x,y,t,\lambda)) d\lambda,
$$
$$
J_{12}=\left(\int\limits_{-\frac{1}{(\Delta t)^{3/4}}}^{-\alpha\sqrt{|x|}}+
\int\limits_{\alpha\sqrt{|x|}}^{\frac{1}{(\Delta t)^{3/4}}}\right) 
\left(\partial_{\tau}\chi_{-I}(x-\lambda y -\lambda^2 t + \omega(x,y,t,\lambda),\lambda)
(-\lambda^2 +\omega_{t}(x,y,t,\lambda)) d\lambda\right),
$$
where we assume $\alpha=1/\sqrt{2(t+1)}$.

Assuming that $|x|\gg1$ and $|x|>\frac{2\left(|y|+ 2 D_y \right)^2}{t+1}$ and using (\ref{eq:3.2}), we immediately obtain:
$$
J_{11}=  O\left(\frac{1}{|x|}\right), \ \ \int\limits_{-\alpha\sqrt{|x|}}^{\alpha\sqrt{|x|}} 
\partial_{\tau}\chi_{L-I}(x-\lambda y -\lambda^2 t,\lambda)
(-\lambda^2) d\lambda = O\left(\frac{1}{|x|}\right);
$$
therefore the exact integral over the interval $[-\alpha\sqrt{|x|},\alpha\sqrt{|x|}]$, as well as the leading term integral over
the same interval are small for large $|x|$. It follows that the exact integral can be replaced by the leading term integral.

From (\ref{eq:5}) we obtain the following estimate of the error in the calculation of $J_{12}$ arising from the 
replacing $\chi_{-I}$ by $\chi_{L-I}$:
$$
\left(\int\limits_{-\frac{1}{(\Delta t)^{3/4}}}^{-\alpha\sqrt{|x|}}+\int\limits_{\alpha\sqrt{|x|}}^{\frac{1}{(\Delta t)^{3/4}}}\right) 
\bigg(\partial_{\tau}\chi_{-I}(x -\lambda y -\lambda^2 t + \omega(x,y,t,\lambda),\lambda) -
$$
$$
 -\partial_{\tau}\chi_{L-I}(x-\lambda y -\lambda^2 t + \omega(x,y,t,\lambda),\lambda) 
(-\lambda^2 +\omega_{t}(x,y,t,\lambda)) d\lambda\bigg)=
$$
$$
=O\left( \int\limits_{\alpha\sqrt{|x|}}^{\frac{1}{(\Delta t)^{3/4}}} \frac{d\lambda}{\lambda^2} \right)=
 O\left(\frac{1}{\sqrt{|x|}}\right)
$$

We proved the following: for $|x|\rightarrow\infty$ the use of $\chi_{L-I}$ instead of $\chi_{-I}$ results into an  
$O\left(\frac{1}{\sqrt{|x|}}\right)$ correction for the $t$-derivative.
Consider now the integrals
$$
J^{(1)}_{12}=\left(\int\limits_{-\frac{1}{(\Delta t)^{3/4}}}^{-\alpha\sqrt{|x|}}+\int\limits_{\alpha\sqrt{|x|}}^{\frac{1}{(\Delta t)^{3/4}}}\right) 
\left(\partial_{\tau}\chi_{L-I}(x-\lambda y -\lambda^2 t+ \omega(x,y,t,\lambda),\lambda)
(-\lambda^2 +\omega_{t}(x,y,t,\lambda)) d\lambda\right)
$$
$$
J^{(1)}_{2}=-\frac 1{\pi}\!\!\! \int\limits_{|\lambda|>\frac{1}{(\Delta t)^{3/4}}} \!\!\! \left[ \chi_{L-I}(x-\lambda y -\lambda^2t
-\lambda^2\Delta t+\omega(x,y,t+\Delta t,\lambda),\lambda) - \chi_{L-I}(x-\lambda y -\lambda^2 t+\omega(x,y,t,\lambda),\lambda) \right]d\lambda
$$
We have $\|\omega_{t}(x,y,t,\lambda)\|_{L^2(d\lambda)}\le{\cal C}_2$; therefore
$$
J^{(1)}_{12}=J^{(2)}_{12}+J^{(2)}_{13}
$$
$$
J^{(2)}_{12}=\left(\int\limits_{-\frac{1}{(\Delta t)^{3/4}}}^{-\alpha\sqrt{|x|}}+\int\limits_{\alpha\sqrt{|x|}}^{\frac{1}{(\Delta t)^{3/4}}}\right) 
\left(\partial_{\tau}\chi_{L-I}(x-\lambda y -\lambda^2 t+ \omega(x,y,t,\lambda),\lambda)
(-\lambda^2) d\lambda\right),
$$
$$
J^{(2)}_{13}=\left(\int\limits_{-\frac{1}{(\Delta t)^{3/4}}}^{-\alpha\sqrt{|x|}}+\int\limits_{\alpha\sqrt{|x|}}^{\frac{1}{(\Delta t)^{3/4}}}\right) 
\left(\partial_{\tau}\chi_{L-I}(x-\lambda y -\lambda^2 t+ \omega(x,y,t,\lambda),\lambda)\,
\omega_{t}(x,y,t,\lambda) d\lambda\right)=O\left(\frac{1}{|x|} \right) ,
$$
Let us estimate the correction to $J^{(2)}_{12}$ due to the term $\omega$ in the argument of $\partial_{\tau}\chi_{L-I}$:
$$
\left|\left(\int\limits_{-\frac{1}{(\Delta t)^{3/4}}}^{-\alpha\sqrt{|x|}}+\int\limits_{\alpha\sqrt{|x|}}^{\frac{1}{(\Delta t)^{3/4}}}\right) \left(\partial_{\tau}[\chi_{L-I}(x-\lambda y -\lambda^2 t + \omega(x,y,t,\lambda),\lambda)-\chi_{L-I}(x-\lambda y -\lambda^2 t,\lambda)]\cdot (-\lambda^2) d\lambda\right)\right| \le
$$
$$
\le 2 \int\limits_{\alpha\sqrt{|x|}}^{\frac{1}{(\Delta t)^{3/4}}} \max\limits_{\tau}|\partial^2_{\tau}\chi_{L-I}(\tau,\lambda)|
\max|\omega|\, \lambda^2\le 2 C \int\limits_{\alpha\sqrt{|x|}}^{\frac{1}{(\Delta t)^{3/4}}}
\frac{\lambda^2 d\lambda}{1+|\lambda^4|}\le \frac{2 C}{\sqrt{|x|}}  
$$
To finish, we have to estimate the corrections in $J^{(1)}_{2}$ due to the $\omega$-term in the argument. Let us use the following splitting:
$$
J^{(1)}_{2}=J^{(2)}_{2}+J^{(2)}_{21}+J^{(2)}_{22},
$$
$$
J^{(2)}_{2}=-\frac 1{\pi}\int\limits_{|\lambda|>\frac{1}{(\Delta t)^{3/4}}} \left[\chi_{L-I}(x-\lambda y -\lambda^2 t-\lambda^2\Delta t,\lambda) - \chi_{L-I}(x-\lambda y -\lambda^2 t,\lambda) \right]d\lambda,
$$
$$
J^{(2)}_{21}=-\frac 1{\pi}\int\limits_{|\lambda|>\frac{1}{(\Delta t)^{3/4}}} \left[\chi_{L-I}(x-\lambda y -\lambda^2 t-\lambda^2\Delta t+\omega(x,y,t+\Delta t,\lambda),\lambda) - \chi_{L-I}(x -\lambda y -\lambda^2 t-\lambda^2\Delta t,\lambda) \right]d\lambda,
$$
$$
J^{(2)}_{22}=\frac 1{\pi}\int\limits_{|\lambda|>\frac{1}{(\Delta t)^{3/4}}} \left[\chi_{L-I}(x-\lambda y -\lambda^2 t+\omega(x,y,t,\lambda),\lambda) - \chi_{L-I}(x-\lambda y -\lambda^2 t,\lambda) \right]d\lambda.
$$
$$
|J^{(2)}_{21}|\le \frac 1{\pi}\int\limits_{|\lambda|>\frac{1}{(\Delta t)^{3/4}}} \max\limits_{\tau}|\partial_{\tau}\chi_{L-I}(\tau,\lambda)|\max|\omega|\le {\cal C}_1 
\int\limits_{\frac{1}{(\Delta t)^{3/4}}}^{\infty} \frac{d\lambda}{|\lambda^3|} = O\left(\Delta t\sqrt{\Delta t}\right).  
$$
Analogously,
$$
|J^{(2)}_{22}|=O\left(\Delta t\sqrt{\Delta t}\right).
$$
Summarizing our calculations from this Section, we obtain the following estimate for fixed $y$, $t\ge0$ and $|x|\rightarrow\infty$:
$$
\partial_t v(x,y,t)- \partial_t v_{L}(x,y,t)=O\left(\frac{1}{\sqrt{|x|}} \right).
$$
From the Pavlov equation in the non-evolutionary form (\ref{Pavlov}), we see that 
$\partial_t v(x,y,t)\big|_{t=0+}$ is constant in $x$ in both intervals $x<-D_x$ and $x>D_x$ outside the support of 
the Cauchy data. Taking $|x|\rightarrow\infty$, we immediately obtain that:
$$
\partial_t v(x,y,t)\big|_{t=0+}=\left\{\begin{array}{ll}
\int_{-\infty}^{+\infty} [v_{y}(x',y,0) +(v_{x'}(x',y,0))^2]_y\,dx', & x<-D_x,\\
0, & x>D_x.
\end{array}\right.
$$
which is consistent only with the choice $\partial_x^{-1}=-\int_x^{\infty}dx'$  (see (\ref{Pavlov-ev2})).
We also obtain that
$$
\partial_t v(x,y,t)\rightarrow 0, \ \ \mbox{for} \ \ x\rightarrow\pm\infty, \ \ t>0,
$$
and, together with the fact that both $\partial_x v(x,y,t), \partial_x v(x,y,t)\rightarrow 0$ for 
$x\rightarrow\pm\infty$, $t>0$, equation (\ref{Pavlov-ev2}) immediately implies the constraint (\ref{eq:man-cond}).
We remark that, once the constraint (\ref{eq:man-cond}) is satisfied, for $t>0$, all possible choices of $\partial_x^{-1}$
become equivalent.

\end{document}